\documentstyle{l-aa}

\input{epsf.sty} 
\newcommand{\bc}{\begin{center}}
\newcommand{\ec}{\end{center}}

\newbox\grsign \setbox\grsign=\hbox{$>$} \newdimen\grdimen \grdimen=\ht\grsign
\newbox\simlessbox \newbox\simgreatbox \newbox\simpropbox
\setbox\simgreatbox=\hbox{\raise.5ex\hbox{$>$}\llap
     {\lower.5ex\hbox{$\sim$}}}\ht1=\grdimen\dp1=0pt
\setbox\simlessbox=\hbox{\raise.5ex\hbox{$<$}\llap
     {\lower.5ex\hbox{$\sim$}}}\ht2=\grdimen\dp2=0pt
\setbox\simpropbox=\hbox{\raise.5ex\hbox{$\propto$}\llap
     {\lower.5ex\hbox{$\sim$}}}\ht2=\grdimen\dp2=0pt

\def\a218{\alpha_{2-18}}
\def\aiso{\alpha_{\rm iso}}
\def\aaniso{\alpha_{\rm aniso}}
\def\alphint{\alpha_{\rm int}}
\def\aint218{\alpha_{{\rm int},2-18}}
\def\elldiss{\ell_{\rm diss}}
\def\eps{\epsilon}

\def\Ldiss{L_{\rm diss}}
\def\Lh{L_h}
\def\Ldiss{L_{\rm diss}}
\def\Ls{L_s}

\def\sigmat{\sigma_{\rm T}}
\def\taut{\tau_{\rm T}}
\def\Te{T_{\rm e}}
\def\Tbb{T_{\rm bb}} 
 
\begin{document}

   \thesaurus{03         
              (02.01.2;  
               02.18.7;  
               02.19.2;  
               11.19.1;  
               13.07.3;  
               13.25.2)} 

   \title{Models of X-ray and $\gamma$-ray emission from Seyfert galaxies}


   \author{Roland Svensson
          }


   \institute{Stockholm Observatory, S--133 36 Saltsj\"obaden, Sweden \\
              Internet: svensson@astro.su.se
             }

   \date{Received December 15, 1995; accepted February 4, 1996}

   \maketitle

   \begin{abstract}

X-ray and $\gamma$-ray observations of Seyfert 1 galaxies are
briefly reviewed.
Both thermal and non-thermal model for X-ray and $\gamma$-ray
emission are discussed.
Particular attention is given to various disc-corona models
including both homogeneous and inhomogeneous (patchy) corona models.
Recent work on exact radiative transfer in such
geometries are reviewed. 


   \end{abstract}

%


\section{X-ray and $\gamma$-ray spectra of Seyfert 1 galaxies}
\label{sec:observations}

\begin{figure}[tb]
\leavevmode
\epsfxsize=8.8cm  \epsfbox{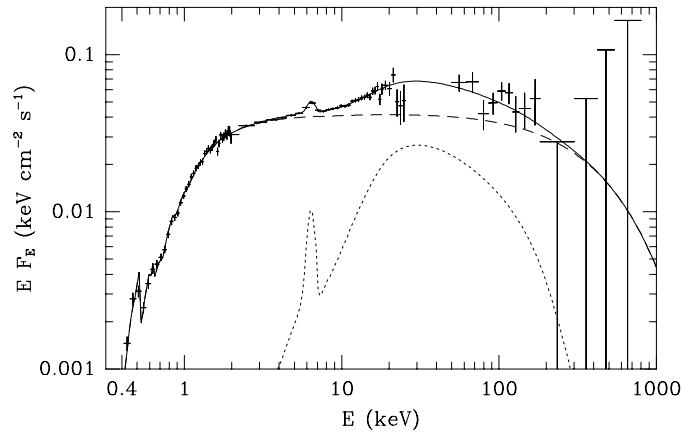}
      \caption{ 
The broad band spectrum (crosses) of IC 4329A from {\it ROSAT},
{\it Ginga}, and OSSE (Madejski et al. 1995). The dashed curve shows the
intrinsic spectrum incident on the cold matter, the dotted curve is the
reflected component, and the solid curve is the sum. From Magdziarz \&
Zdziarski (1995).
              }
         \label{fig1}
   \end{figure}

By combining observations in different energy bands, a picture has emerged 
where the overall shapes of the X-ray and $\gamma$-ray spectra of
different Seyfert 1 galaxies are similar.
The only Seyfert 1 galaxy for which the broad band X- and $\gamma$-ray
spectrum is well determined is IC 4329A (Madejski et al. 1995). 
 Combined {\it ROSAT}, {Ginga} and OSSE data are shown in 
 Fig.~\ref{fig1}. The observations were modelled by 
Madejski et al. (1995), Zdziarski et al. (1994), and  
Magdziarz \& Zdziarski (1995). The spectrum consists of two components:
1) an intrinsic power law  of slope $\alphint 
\approx 0.9$ with an exponential cutoff
energy of about 300 keV (dashed curve in  Fig.~\ref{fig1}), 
and 2) a reflection component caused by 
cold reflecting matter subtending a solid angle $\sim 1-2 \pi$ 
as viewed from the X-ray source (dotted curve in  Fig.~\ref{fig1}).
The reflection component peaks at about 30 keV 
causing the overall 2-18 keV
spectral slope to be considerably flatter, $\a218 \approx 0.7$ and
the apparent cutoff energy to be smaller, $\approx 50$ keV.

The average X/$\gamma$-spectrum for four Seyfert 1s using nonsimultaneous
 {\it Ginga} and OSSE data (23 {\it Ginga} spectra and 8 OSSE spectra)
shows similar properties as the
spectrum for IC 4329A (Zdziarski et al. 1995).
The same holds for the nonsimultaneous average {\it EXOSAT}/OSSE
spectrum for seven Seyfert 1s using 31 {\it EXOSAT} and 18
OSSE observations (Gondek et al. 1996). 
These spectra are consistent with the 
average spectrum of these Seyfert 1s from {\it HEAO-1} A4 (Gondek et al. 1996)
and with the upper limits of all Seyferts from COMPTEL 
(Maisack et al. 1995).

Larger samples exist within narrower spectral ranges using
a single experiment. These samples show spectra consistent
with the smaller samples above.
For example, 60 {\it Ginga} spectra of 27 Seyferts
show $\aint218 \approx 0.95$ and the overall $\a218 \approx 0.73$,
each having a dispersion of about 0.15 (Nandra \& Pounds 1994).

The brightest Seyfert 1 (or 1.5), NGC 4151, once considered a prototype
for Seyfert 1s, has now turned out to have unusual X-ray
properties. In the 2-18 keV {\it Ginga}-range, the spectral index, 
$\a218$, varies in the range 0.3-0.7 instead of having the
canonical Seyfert value of 0.7. The harder the spectrum is,
the weaker is the 2-18 keV flux, and the spectrum seems to pivot around
100 keV. And indeed, OSSE observations during 5 viewing periods from
1991-1994 show that the OSSE flux is essentially constant (25 \% day-to-day
variability; Johnson et al 1996). Furthermore, the OSSE observations 
show that the spectrum cuts off at about 50 keV, much less
than the canonical 300 keV value. 
Finally, there seems to be no reflection component.
NGC 4151 is clearly a freak object
which requires special consideration to fit into the unified scheme
of Seyfert galaxies.

\section{Outstanding questions regarding the X-ray emitting regions
in Seyfert 1 galaxies}
\label{sec:outstanding}

The obvious question is of course which 
radiation process generates the X-ray continuum.
It is commonly believed that Compton scattering by
energetic electrons, either mildly relativistic thermal electrons
or highly relativistic nonthermal electrons, Compton
scatter soft UV-photons into the X-ray range.

Then there is the question of geometry, i.e. the
spatial distribution 
of hot X-ray generating electrons and of cold 
($T < 10^6$ K) reflecting matter. Possible geometries
include a cold slab surrounded either by plane parallel
coronal slabs of hot electrons, or by coronal patches (active regions)
of unknown geometries.

Finally, there is the question whether it is thermal or nonthermal
electrons or both that account for the X-ray emission.
This question might be answered if the spectrum above a few 
hundred keV was known with certainty, but the best available
spectrum of the brightest typical Seyfert 1, IC 4329A, does not have 
sufficient signal to noise at such energies.
The spectrum of IC 4329A can be fit with both thermal
and nonthermal models (Zdziarski et al. 1994).

The nonthermal models studied in the 1980s 
have several attractive features (for review,
see Svensson 1994). In particular, it was predicted already
in 1985 (Svensson 1986) that the pair cascades in the
nonthermal models give rise to an X-ray spectrum 
with $\aint218 \approx 0.9-1.0$ in contradiction to the
overall slope of $\a218 \approx 0.7$ but in agreement
with the intrinsic spectral spectral slope
later found by {\it Ginga} (e.g. Nandra \& Pounds 1994).

Due to the non-detections by COMPTEL and the high energy
cutoffs indicated by OSSE, attention has recently been
focused on the thermal models.
 

\section{Geometry of the X-ray source}
\label{sec:geometry}

In the unified model for Seyfert galaxies (e.g. Antonucci 1993), 
we are directly viewing the central X-ray source and broad
line region in Seyfert 1 galaxies, but our line-of-sight
passes through an optically opaque molecular torus in 
Seyfert 2 galaxies. In standard accretion disk scenarios
for the central X-ray source, the unified models imply
that our viewing angle (i.e. the angle between
the disk normal and the line-of-sight) is less than the
half-opening angle of the ionization cone. Half-opening angles  inferred
from observations are typically 30-40 degrees, which means that
in Seyfert 1s we are viewing the accretion disk from directions
that are closer to face-on than to edge-on.

In principle, the central X-ray source could have spherical symmetry
with the source of UV-photons being small cold clouds
uniformly distributed throughout the X-ray source.
In this case, radiation models invoking spherical symmetry
would be applicable (e.g., 
Ghisellini \& Haardt 1994, Skibo et al. 1995; Pietrini \& Krolik 1995).
In some models, the central X-ray source is quasi-spherical but the
UV-source is anisotropic. Here, the X-ray source can either be a hot
two-temperature accretion disk in local thermal balance
(Shapiro, Lightman, \& Eardley 1976) or a two-temperature ion
torus (Rees et al. 1982, recently rejuvenated as advection dominated
accretion flows, e.g. Abramowicz et al. 1995, where the connections 
between the two types of hot disks with other disk solutions are 
elucidated). The UV-source may then be the standard cold 
Shakura-Sunyaev disk located beyond some transition radius.
This anisotropy of the UV-photons breaks the spherical symmetry.
Then detailed radiative transfer calculations of
how the UV-photons penetrate the hot disk are required in order 
to determine the emerging X-ray spectrum of Comptonized UV-photons 
at different viewing angles. 
This problem has not yet been solved.

The most commonly used scenario is the two-phase disk-corona model
(e.g., Haardt \& Maraschi 1991, 1993) in which a hot X-ray emitting corona 
is located above the cold UV-emitting disk of the canonical 
black hole model for AGNs. The power law X-ray spectrum
is generated by thermal Comptonization of the soft UV-radiation.
About half of the X-rays enters and is reprocessed by the cold disk,
emerging mostly as black body disk radiation in the UV.
Haardt \& Maraschi (1991) emphasized the coupling between the 
two phases due to the reprocessing, as the soft disk photons
influence the cooling of the corona. They showed that nearly all
power must be dissipated in the corona in order to have
$\alphint \sim 0.9-1$.
 A consequence of this is that the soft disk
luminosity, $\Ls$, is of the same order as the hard X-ray luminosity,
$\Lh$. The disk-corona scenario is highly anisotropic as the
UV-photons enter the corona from below only.

Observations show that $\Ls$ may be several times larger than 
$\Lh$, in contradiction to the prediction of the uniform two-phase
disk-corona model. This led Haardt, Maraschi, \& Ghisellini (1994)
to propose a patchy disk-corona model, where the corona consists 
of several localized active regions on the surface of the disk.
Internal disk dissipation results in UV-radiation that leaves
the disk without entering the active regions, thus enhancing 
the observed $\Ls$ relative to $\Lh$. The patchy corona model
is also highly anisotropic as the UV-radiation enters the
active regions from below and through the sides.

The   full radiative
transfer and Comptonization problem in the these geometries
has recently been solved (Haardt \& Maraschi 1993, Stern et al.1995b,
Poutanen \& Svensson 1996). 
We now turn to discuss the methods and the results.


\section{Radiative transfer/Comptonization} \label{sec:transfer}

Two different methods have been used  to solve the full
radiative transfer/Comptonization problem in mildly relativistic
thermal plasmas accounting for energy and pair balance as well
as reprocessing by the cold disk (including angular anisotropy and
Klein-Nishina effects).

The first method is based on the Non-Linear Monte Carlo (NLMC) method
developed by Stern (1985, 1988) and described in detail in
Stern et al (1995a). The Monte Carlo particles (particles and photons)
make up the background so the Monte Carlo particles can interact
with themselves (e.g. photon-photon pair production) causing
nonlinearity. Any geometry can be treated including 2D and 3D
geometries, but calculations so far (Stern et al 1995b)
have been limited to coronal slabs (1D) or active regions (2D) in the
shape of hemispheres or spheres at different elevations
above the cold disc. Another advantage is the possibility to
divide the region into several zones in order to study the
inhomogeneous distributions of, e.g., temperature and pair density.
The drawback is that each run takes a few hours on a Sun IPX.

The second method is based on the iterative scattering method (ISM),
where the radiative transfer is exactly solved for each scattering order
separately (e.g. Sunyaev \& Titarchuk 1985, Haardt 1994).
The code and its testing are briefly described in
Poutanen and Svensson (1996, see also  Poutanen 1994; Poutanen \& Vilhu
1993). It was shown that the results
of the NLMC and the ISM code are in excellent agreement.
The ISM code is a 1D code but it can also treat quasi-2D radiative transfer
in cylinders/pill boxes and hemispheres. The full Compton scattering 
matrix is used allowing to solve polarized radiative transfer
in thermal relativistic plasmas. Fully angular dependent, polarized
Compton reflection is implemented using a Green's matrix 
(Poutanen, Nagendra \& Svensson 1996). The advantage of the ISM code
is that it takes 10 minutes or less on a Sun IPX.

\section{The Comptonization solution} \label{sec:compsolution}

\begin{figure}[htbp]
\bc
\leavevmode
\epsfxsize=10.8cm  \epsfbox{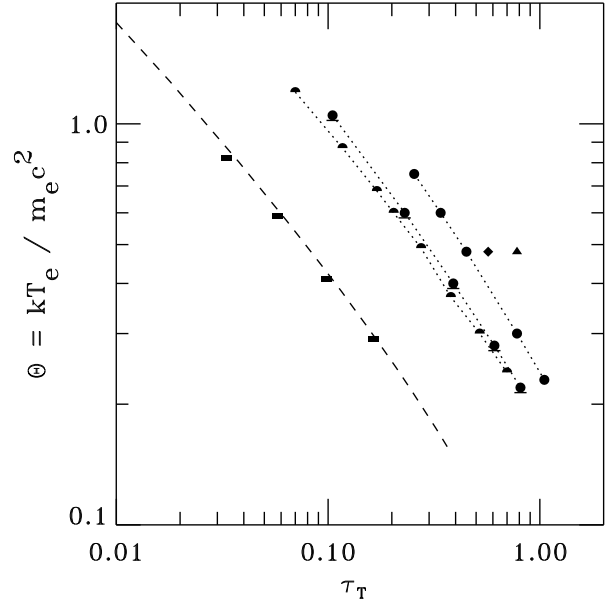}
\ec
\caption{Dimensionless volume-averaged temperature, $\Theta \equiv kT_e/ m_e c^2$,
vs. Thomson scattering optical depth, $\taut$, for
a steady X-ray emitting  plasma region on or above a cold disk surface.
The plasma Compton scatters reprocessed soft black body photons from
the cold disk surface. 
{\it Solid rectangles} and {\it dashed curve} show results
from NLMC code and ISM code, respectively, for
the case of a plane-parallel slab corona.  
Results using the NLMC code for individual active pair regions
are shown for
{\it hemispheres} located on the disk surface;
surface spheres also located on the surface ({\it underlined spheres});
spheres located at a height of 0.5$h$ ({\it spheres}),
1$h$ ({\it diamond}), and 2$h$ ({\it triangle}), where $h$ is the radius
of the sphere.
The results for each type of active region are connected by
{\it dotted} curves. The cold disk temperature,  $k\Tbb$ = 5 eV. 
From Stern et al. (1996).
}
\label{temptaut}
\end{figure}

The two codes have been used to study radiative transfer/Comptonization in pure
pair coronae in energy and pair balance. For coronae of a given geometry
and in energy 
balance, there exists a unique $T_e - \taut$ relation, 
where $\Te$ is the volume-averaged coronal temperature 
and $\taut$ is a characteristic 
Thomson scattering optical depth of the coronal region.
For slabs, $\taut$ is the vertical Thomson scattering optical depth.
For hemispheres and for spheres, the radial $\taut$ is averaged over $2\pi$ solid
angle for hemispheres, and $4\pi$ solid angle for spheres.

In Figure~\ref{temptaut}, this relation is shown for different geometries.
The results for slabs are shown by {\it rectangles}, 
for hemispheres located on the disk surface by {\it hemispheres}, 
for surface spheres located on the disk surface by {\it underlined spheres},
and for spheres located at heights, 0.5$h$, 1$h$, and 2$h$, by {\it spheres},
{\it diamonds}, and {\it triangles}, respectively 
($h$ is the radius of the sphere).
The results for active regions are connected by {\it dotted} curves.
For comparison we also show the slab results from Stern et al. 1995b
 using the ISM code ({\it dashed curve}).
Each curve is characterized by an almost constant generalized
Kompaneets parameter, $y \equiv \taut (1 + \taut) (4 \Theta + 16 \Theta^2)$,
where  $\Theta \equiv kT_e/ m_e c^2$.
The larger the soft photon starvation (i.e. the fewer the number of 
reprocessed soft photons
reentering the coronal region), the larger is $y$. 
There is very good agreement  
between the slab results from the NLMC and the ISM codes,
which tests the accuracy of both codes and methods.

\section{Pair balance and the compactness}
\label{sec:pairbalance}

\begin{figure}[htbp]
\bc
\leavevmode
\epsfxsize=10.cm  \epsfbox{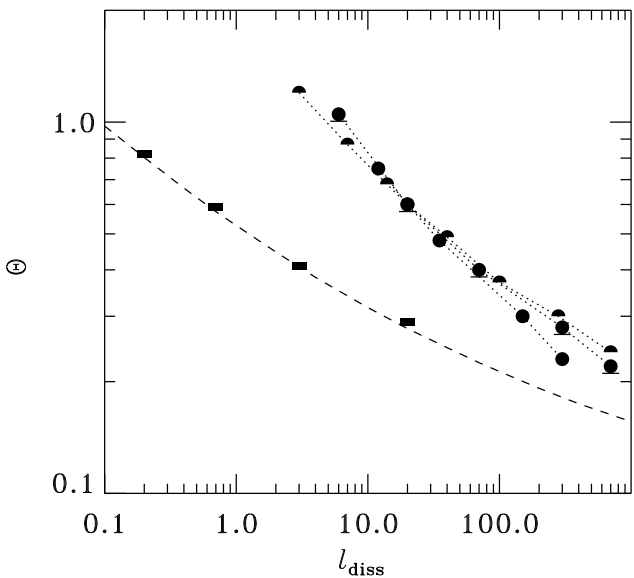}
\ec
\caption{
Dimensionless temperature, $\Theta \equiv kT_e/ m_e c^2$,
vs. dissipation compactness, 
$\elldiss \equiv ( \Ldiss /h )$ $(\sigmat /  m_e c^3)$,
for a steady X-ray emitting pair plasma 
in pair and energy balance located on or above
a cold disk surface.
Same symbols and notation as in Fig.~2.
The parameter space to the right of respective curves is 
forbidden as pair balance cannot be achieved, and  the parameter space
to the left would contain solutions where the background 
coronal plasma dominates over the pairs, i.e. ``pair free" solutions 
(e.g. Svensson 1984, HM93). 
The soft photons from the cold disk are assumed to
have $k\Tbb$ = 5 eV. 
From Stern et al. (1996).
}
\label{tempelldiss}
\end{figure}

Solving the pair balance for the combinations of ($\Theta$, $\taut$)
obtained in \S~\ref{sec:compsolution} gives a unique dissipation
compactness, $\elldiss$ (see Ghisellini \& Haardt 1994 for a discussion).
Here, the local dissipation compactness,
$\elldiss \equiv ( \Ldiss /h ) (\sigmat /  m_e c^3)$,
characterizes the dissipation with $\Ldiss$ being the power providing uniform
heating in a cubic volume of size $h$ in the case of a slab of height $h$, 
or in the whole
volume in the case of an active region of size $h$.
Figures~\ref{tempelldiss}  shows the 
$\Theta$ vs. $\elldiss$ relation obtained with the NLMC code
for different geometries and for $k\Tbb$ = 5 eV.
Notations are the same as in Figure~\ref{temptaut}. 
The slab results with the ISM codes are also shown by the dashed curve.  
The parameter space to the right of respective curve is 
forbidden as pair balance cannot be achieved, and  the parameter space
to the left would contain solutions where the background 
coronal plasma dominates over the pairs, i.e. ``pair free'' solutions 
(e.g. Svensson 1984, HM93). 

From Figure~\ref{tempelldiss} we find that at a given $\Theta$
the ARs have a larger $\elldiss$ than the slabs.
This is due to the longer escape times in slabs.
In a local cubic volume in slabs, the four vertical surfaces act
effectively as ``reflecting'' surfaces increasing the radiation 
energy density in the volume
as compared to the ARs where the radiation escapes freely through
the surface of the AR volume.

\section{Anisotropy effects}
\label{sec:anisotropy}

\begin{figure}[htbp]
\bc
\leavevmode
\epsfxsize=10.cm  \epsfbox{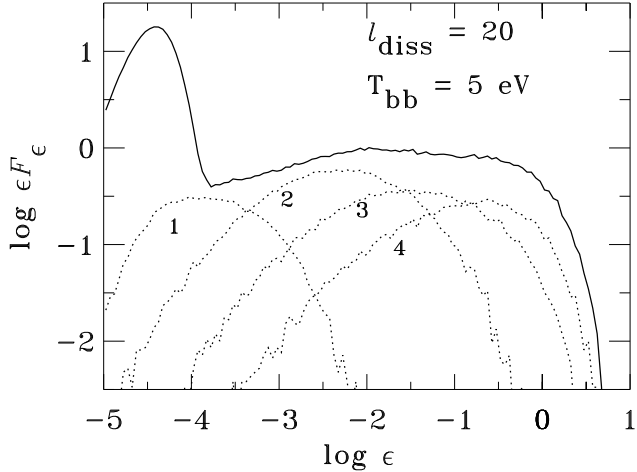}
\ec
\caption{Emerging spectrum, $\eps F_{\eps}$, where $F_{\eps}$ is the energy
flux (arbitrary units) and $\eps \equiv h \nu / m_e c^2$ from hemisphere
with compactness, $\elldiss$ = 20, and cold disk temperature, $k \Tbb$ = 5 eV.  
The spectrum is averaged over
viewing angles $0.6 < \cos \theta < 1$.
The {\it  solid curve} shows the sum of the Comptonized spectrum 
from the hemisphere and the reprocessed
black body spectrum (the reflection component is not included).
The numbered {\it dotted curves} show the different scattering orders. 
Note that the first scattering order is significantly reduced due to anisotropic
Compton scattering. This deficiency causes an {\it anisotropy break} close to the
peak of the second scattering order. Below the anisotropy break, the spectrum is
a power law with an {\it anisotropy slope} that is harder than the standard
{\it isotropy slope} above the break.
From Stern et al. (1996).
}
\label{spectrum}
\end{figure}

Figure~\ref{spectrum} shows the the  ``face-on'' emerging spectrum for a hemisphere
with $\elldiss$ = 20 and $k\Tbb$ = 5 eV with the reflected component subtracted.
The {\it solid curve} is the sum of the reprocessed soft black body spectrum and the
Comptonized spectrum. The ``face-on'' spectrum is averaged over 
viewing angles $0.6 < \cos \theta < 1$.
The numbered {\it dotted curves} show spectral profiles of
the different scattering orders. As noted by Haardt (1993) and HM93, the  
first scattering order is significantly reduced in face-on directions due to anisotropic
Compton scattering when $\Theta$ is mildly relativistic. 
This deficiency at low energies causes an {\it anisotropy break} close to the
peak of the second scattering order. Below the anisotropy break, the spectrum is
a power law with an {\it anisotropy slope}, $\aaniso$, that is harder than the standard
{\it isotropy slope}, $\aiso$, above the break. It is the standard isotropy slope, $\aiso$,  
that  has been fitted with several analytical expressions that are functions
of $\Theta$ and $\taut$ (e.g., Zdziarski 1985, Titarchuk 1994). ``Edge-on''
spectra have a much weaker reflection component and no anisotropy break.

\section{Spectra from active pair regions: hemispheres}
\label{sec:hemispectra}

\begin{figure}[htbp]
\bc
\leavevmode
\epsfxsize=10.cm  \epsfbox{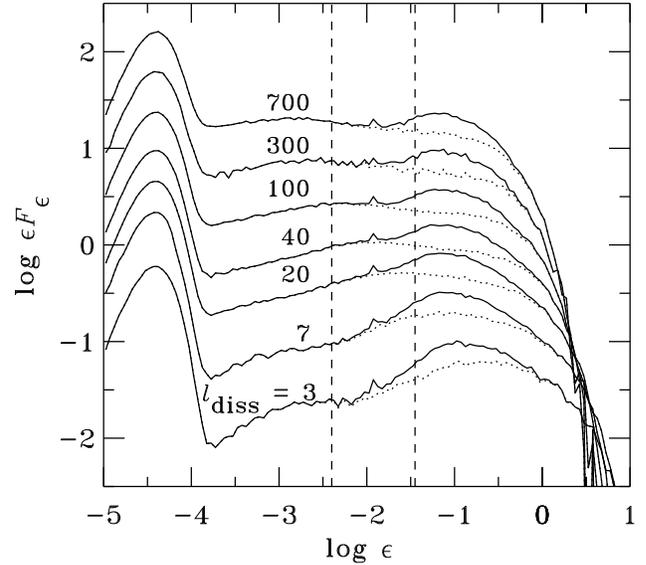}
\ec
\caption{Emerging ``face-on'' spectra, $\eps F_{\eps}$, 
where $F_{\eps}$ is the energy
flux (arbitrary units) and $\eps \equiv h \nu / m_e c^2$, from hemispheres
of different compactnesses, $\elldiss$ and for cold disk temperature, $k\Tbb$ = 5 eV. 
The  ``face-on'' spectra are averaged over viewing angles $0.6 < \cos \theta < 1$.
The {\it  solid curves} show the total spectrum which is the
sum of the Comptonized spectrum from the hemisphere itself ({\it dotted curves}),
the reprocessed black body spectrum, and the reflection component.
Vertical {\it dashed lines} show the 2-18 keV spectral range. 
The anisotropy break in the face-on spectra moves to lower photon energies 
as $\elldiss$ increases.
From Stern et al. (1995b).
} 
\label{hemispectra}
\end{figure}

Figure~\ref{hemispectra} shows emerging ``face-one'' spectra
from hemispheres at different $\elldiss$
for $k\Tbb$ = 5 eV.
We have $\aaniso \approx$ 0.77 
while $\aiso \approx$ 1.09. The anisotropy break is therefore about 0.3.
The anisotropy break moves through the 2-18 keV range for $\elldiss \approx$ 10-100,
and thus $\a218$ is smaller than unity for $\elldiss$ less than 100. 
At this low $k\Tbb$ the first Compton order hardly extends into the 2-18 keV range
and there is little  discreteness effect on the behaviour of $\a218$.

\section{Diagnostics Using Compactness and {\it Ginga} Slopes}
\label{sec:compginga}

\begin{figure}[htbp]
\bc
\leavevmode
\epsfxsize=10.3cm  \epsfbox{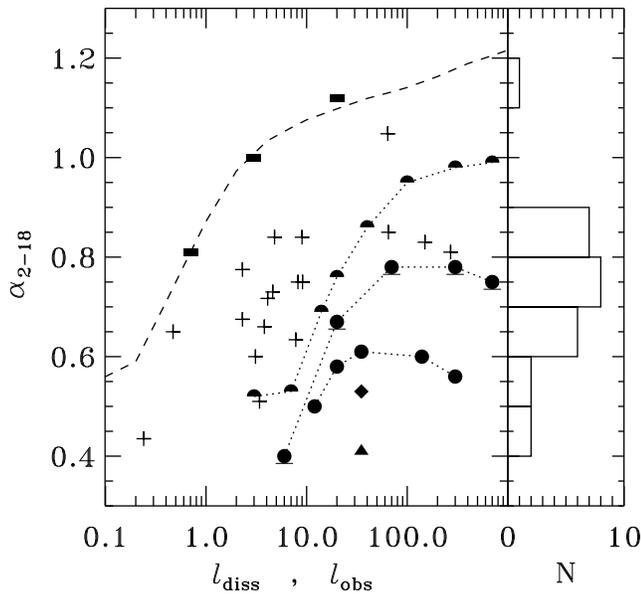}
\ec
\caption{Overall spectral intensity index, $\a218$, least square fitted 
to the model spectra in the 2-18 keV range vs. the dissipation compactness,
$\elldiss$. Same notation as in Fig.~2.
The spectra from NLMC code were averaged over viewing  angles
$0.6 < \cos \theta < 1.0$ before determining $\a218$ with least square fitting.
For the ISM code, face-on spectra (at $\cos \theta = 0.887$) were used.
The right panel shows the observed distribution of
$\a218$ for {\it Ginga} spectra from 27 Seyfert galaxies (Nandra \& Pounds 1994). 
The crosses represent 17 Seyfert galaxies that have both been observed by 
{\it Ginga}
and have known estimates of their X-ray time variability,
$\Delta t$, and thus of the observed compactness,
$\ell_{\rm obs} \equiv ( L_{\rm obs} /c \Delta t ) (\sigmat /  m_e c^3)$
.}
\label{alphaell5ev} 
\end{figure}

The least square overall spectral slope, $\a218$, for the
2-18 keV range were determined and are displayed
in Figure~\ref{alphaell5ev} as a function of the dissipation compactness,
$\elldiss$, for different geometries.
The right panel of Figure~\ref{alphaell5ev} shows the observed distribution of
$\a218$ for {\it Ginga} spectra from 27 Seyfert galaxies (Nandra \& Pounds
1994). The crosses represent 17 Seyfert galaxies that have both been 
observed by {\it Ginga}
and have known estimates of their X-ray time variability 
(and thus lower limits of their compactnesses). The true crosses may 
lie to the right of the plotted ones. 

One sees that the observations are more consistent with active surface regions
(such as hemispheres) 
than with slabs.  Active regions produce spectra covering
the {\it observed ranges} of $\a218$ ($\approx 0.4-0.9$)  and 
cutoff energies ($\sim 2kT_e$)
for the {\it observed range} of compactnesses.

%
%

\section{Conclusions}

Have we been able to throw some light upon the outstanding questions
discussed in \S~\ref{sec:outstanding}?

Regarding the radiative transfer/Comptonization, the progress has been rapid and
it is now possible to use full radiative transfer simulations when fitting
X-ray observations, e.g. with XSPEC. 
As anisotropic effects are very important, i.e. as 
spectral shapes depend strongly on viewing angle, it will be possible
to set constraints on the viewing angle.

The spectra also depend on the geometry of the coronal regions, so
observed spectra can be used as diagnostics of the geometry.
Presently, it seems that active regions are favoured over homogeneous
slab coronae.

There is still no answer to the question whether the radiating electrons
(and positrons) are thermal or nonthermal. High quality spectra above 200 keV
are needed, and the question may not be settled before the 
launch of {\it INTEGRAL}.
    
\begin{acknowledgements}
      The author thanks the Swedish Natural Science Research Council,
      which fully supported (including the registration fee)
      the invited participation in the 3rd Compton Symposium.
      He also thanks Juri Poutanen for valuable comments.
\end{acknowledgements}

\end{document}